\documentclass[runningheads]{llncs}
\usepackage[T1]{fontenc}
\usepackage{graphicx}
\usepackage{lineno,hyperref}
\usepackage{booktabs}
\usepackage{multirow}
\usepackage{xcolor}
\usepackage{hyperref}
\usepackage{marvosym}

\begin{document}
\title{WavFusion: Towards wav2vec 2.0 Multimodal Speech Emotion Recognition}
%
%
\author{Feng Li\inst{1,2}\textsuperscript{(\Letter)} \and
Jiusong Luo\inst{1} \and
Wanjun Xia\inst{1}}
\authorrunning{F. Li et al.}
%
\institute{Department of Computer Science and Technology, Anhui University of Finance and Economics, Anhui, China
\\
 \and
School of Information Science and Technology, University of Science and Technology of China, Anhui, China\\
\email{lifeng}@aufe.edu.cn}
\maketitle              
\begin{abstract}
Speech emotion recognition (SER) remains a challenging yet crucial task due to the inherent complexity and diversity of human emotions. To address this problem, researchers attempt to fuse information from other modalities via multimodal learning. However, existing multimodal fusion techniques often overlook the intricacies of cross-modal interactions, resulting in suboptimal feature representations. In this paper, we propose  WavFusion, a multimodal speech emotion recognition framework that addresses critical research problems in effective multimodal fusion, heterogeneity among modalities, and discriminative representation learning. By leveraging a gated cross-modal attention mechanism and multimodal homogeneous feature discrepancy learning, WavFusion demonstrates improved performance over existing state-of-the-art methods on benchmark datasets. Our work highlights the importance of capturing nuanced cross-modal interactions and learning discriminative representations for accurate multimodal SER. Experimental results on two benchmark datasets (IEMOCAP and MELD) demonstrate that WavFusion succeeds over the  state-of-the-art strategies on emotion recognition.

\keywords{Speech emotion recognition \and multimodal \and wav2vec 2.0 \and A-GRU \and A-GRU-LVC.}
\end{abstract}
\section{Introduction}

Recently, speech emotion recognition (SER) is a fascinating field that utilizes technology to analyze and identify different emotions present in human speech \cite{el2011survey}. This technology has various applications, including in customer service and market research \cite{li2021speech}, learning and education \cite{li2007speech}, mental health \cite{elsayed2022speech}, and social media analytics \cite{ahire2022emotion}. In real-life scenarios, humans express emotions not only through speech but also through alternative modalities, such as text and visuals \cite{calefato2017emotxt,you2016building}. Previous studies on SER typically rely on speech information. However, different modalities provide complementary information for emotion recognition, and emotion recognition of the single modality is not inadequate to meet real-world demands. To address this problem, researchers utilize multimodal information to identify emotional states \cite{abdullah2021multimodal}. In the domain of Multimodal Emotion Recognition (MER), the information of diverse modalities is complementary, providing additional cues to mitigate semantic and emotional ambiguities.

In addition to multimodality, another challenge of SER is achieving better interaction during the fusion of different modalities. Firstly, multimodal data often exhibit asynchrony \cite{wu2021emotion}. For instance, visual signals typically precede audio signals by approximately 120 ms in emotional expressions  \cite{grant2001speech}. This asynchronicity poses a challenge to feature fusion and model design, necessitating methods to address temporal alignment and matching issues. To address this issue, Tsai et al. \cite{tsai2019multimodal} have proposed specific asynchronous models and cross-modal attention mechanisms. Zheng et al. \cite{zheng2022multi} solved heterogeneity among different encoder output features by employing unsupervised training of a multi-channel weight-sharing autoencoder. This approach minimizes the differences among features extracted from different modalities. Additionally, the interactions are simulated by supervised training of cascaded multi-head attention mechanisms. However, most methods with cross-modal attention mechanisms ignore redundant information during the fusion process, thus restricting the performance of MER. Additionally, samples with the same emotion in multimodal data may exhibit differences across modalities, referred to as homogeneous feature differences. For instance, some features in speech and text may exhibit formal similarity but convey different emotional states \cite{chen2021multimodal}. Hazarika et al.\cite{hazarika2020misa} projected each modality into two different subspaces capturing modality-invariant and modality-specific features. However, they only considered the differences between the different emotion of same modalities and ignored the differences between different modalities with the same emotion.  
DialogueTRM explores intra- and inter-modal emotional behaviors in conversations, using Transformers to model the context \cite{mao2020dialoguetrm}. MMGCN proposes a multimodal fusion approach via a deep graph convolution network, modeling the interactions between different modalities using a graph \cite{hu2021mmgcn}. MM-DFN introduces a dynamic fusion network that leverages intra- and inter-modal information at different levels of representation \cite{hu2022mm}. M2FNet proposes a multi-modal fusion network that learns and fuses complementary information from audio, visual, and textual modalities \cite{chudasama2022m2fnet}.

Therefore, in this paper, we propose a novel arhitecture called WavFusion for emotion recognition. Unlike DialogueTRM, WavFusion specifically focuses on incorporating wav2vec2.0 \cite{baevski2020wav2vec} with a gated cross-modal attention mechanism to dynamically fuse multimodal features. Additionally, WavFusion introduces multimodal homogeneous feature discrepancy learning to distinguish between same-emotion but different-modality representations. WavFusion does not rely on graph-based modeling but instead uses a transformer architecture with a modified cross-modal attention mechanism. WavFusion also emphasizes capturing both global and local visual information through the A-GRU-LVC module. While MM-DFN focuses on dynamic fusion strategies, WavFusion emphasizes the use of wav2vec2.0 pre-trained representations and a gated cross-modal attention mechanism to mitigate redundant information during fusion. Additionally, WavFusion incorporates multimodal homogeneous feature discrepancy learning to distinguish between representations of the same emotion across different modalities.

The main contributions of this paper can be summarized as follows:
\begin{itemize}
	\item We propose a multimodal speech emotion recognition model (WavFusion) that leverages the power of wav2vec 2.0 and incorporates textual and visual modalities to enhance the performance of audio-based emotion recognition. 
	\item We integrate the designed gated cross-modal attention mechanism into the wav2vec 2.0 model to mitigate redundant information during the fusion process. Meanwhile, we employ multimodal homogeneous feature discrepancy learning to enhance the discriminative capability of the model.
	\item Experimental results on two benchmark datasets demonstrate the effectiveness of the proposed method. Our WavFusion succeeds over existing state-of-the-art methods.
\end{itemize}

\section{Proposed Method}

\subsection{Problem Statement}

Given a multimodal signal $S_j=\left\{S_j^a, S_j^t, S_j^v\right\}$, we can represent the unimodal raw sequence extracted from the video fragment $j$ as $S_j^m, m \in\{a, t, v\}$. Here, the modalities are denoted by $\{a, t, v\}$, which refer to audio, text, and visual modalities. 

In WavFusion, we aim to predict the emotion category for each utterance. It focuses on categorizing the emotion conveyed in each utterance, assigning it to a specific emotion class or category, $y_j\in{R}^c$. $c$ is the number of emotion categories. Figure 1 illustrates the overall structure of WavFusion, including an auxiliary modal encoder, a primary modal encoder, and multimodal homogeneous feature discrepancy learning. The orange color represents positive emotions and the green color represents negative emotions.

\subsection{Auxiliary Modality Encoder}

\subsubsection{Video Representation}

For the visual modality, we use the EfficientNet pretrained model as a feature extractor to obtain visual features $\mathbf{e}_j^v$. This model is a self-supervised framework for visual representation learning. In this paper, we attempt to extend EfficientNet to emotion recognition. $\mathbf{e}_j^{\mathrm{v}}$ can be formulated as:

\begin{equation}
	\mathbf{e}_j^v=\Phi_{ {visual}}\left(S_j^v\right)
\end{equation}
where $\Phi_{{visual}}$ denotes the function of EfficientNet model.

\begin{figure*}[t]
	\centering
	\includegraphics[width=1\linewidth]{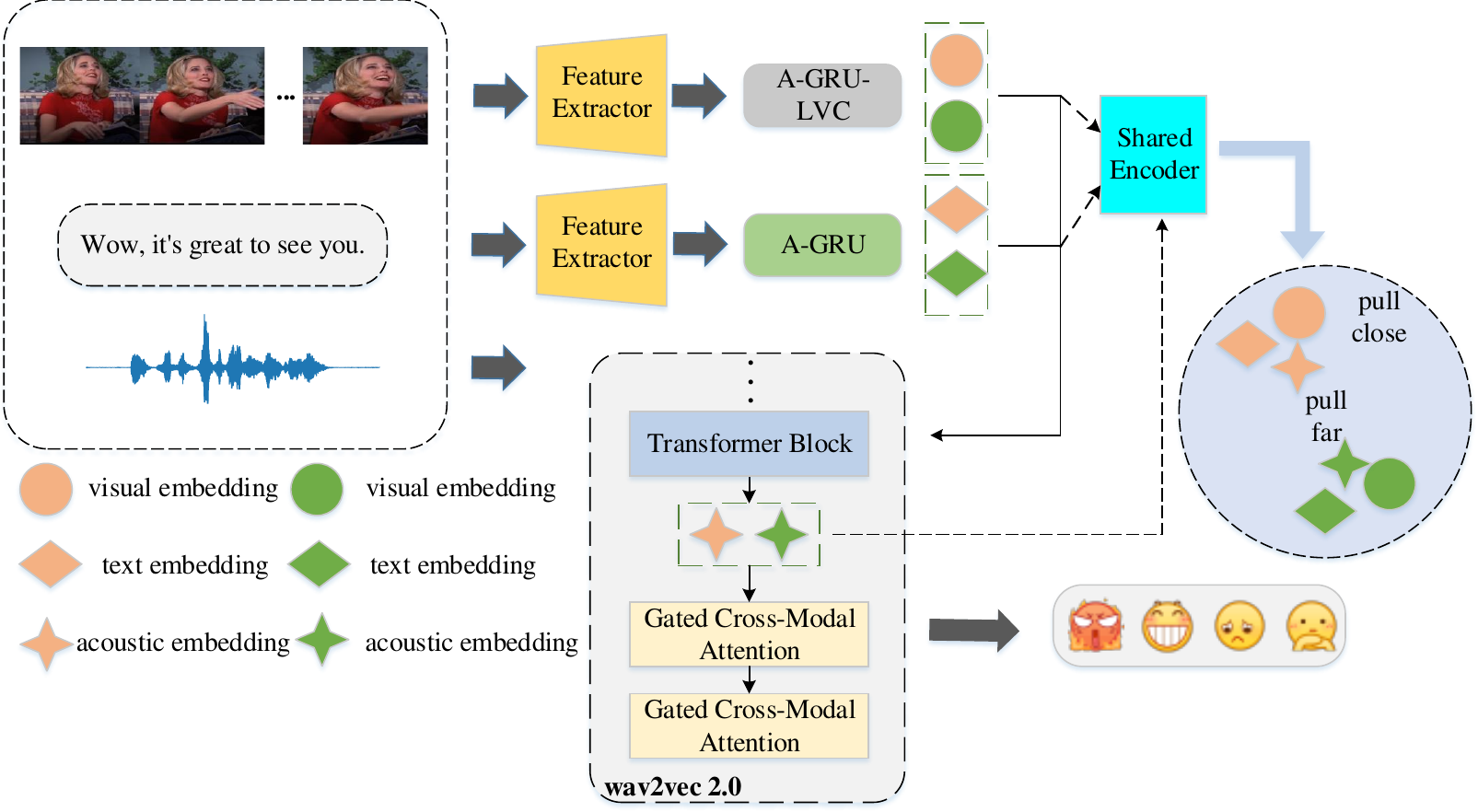}
	\caption{The overview of WavFusion.}
\end{figure*}

On the other hand, we consider the context and situation conveyed by the global information in the visual modality, along with the specific details of actions and expressions from local information. The visual feature is fed into the proposed A-GRU-LVC module, which aims to extract both global and local features. 
\begin{equation}
	X_j^{v 1}=F_{SA}\left(F_{GRU}\left(\mathbf{e}_j^v\right)\right)
\end{equation}
where $F_{SA}$ and $F_{GRU}$ denote the learning functions of GRU and self-attentive mechanism, respectively.

Simultaneously, to preserve local corner point regions and extract local information, a learnable visual center (LVC) is implemented on the visual features \cite{quan2023centralized}. This LVC aggregates features from local areas, ensuring that important local information is retained. In contrast to the approach, we utilize one-dimensional convolution instead of two-dimensional convolution.

\begin{equation}
	X_j^{v2}=F_{LVC}\left(e_j^v\right)
\end{equation}
where $F_{LVC}$ denotes the learning functions of the LVC block.

Finally, the output of the A-GRU-LVC block is obtained by connecting the output of the self-attention module $X_j^{v1}$ and the output of the LVC block $X_j^{v2}$ along the last dimension.
\begin{equation}
	X_j^{v}=X_j^{v1}\oplus X_j^{v2}
\end{equation}

\subsubsection{Contextualized Word Representation}
To capture rich contextual information from textual data, we utilize the RoBERTa-base model, which belongs to the transformer family, as a contextual encoder. The architecture of RoBERTa consists of multiple Transformer layers, including a stack of encoders. Each encoder layer contains a multi-head self-attention mechanism and a feed-forward neural network. RoBERTa is designed to capture contextualized representations of words in a sentence, allowing it to understand the meaning and relationships between different words. 
$\mathbf{e}_j^t$ can be formulated as:

\begin{equation}
	\mathbf{e}_j^t=\Phi_{{text}}\left(S_j^t\right)
\end{equation}
where $\Phi_{{text}}$ denotes the function of the RoBERTa pre-train model.

To further consider context-sensitive dependence for text features, we feed it into the GRU and the self-attention mechanism to obtain global features of the text information. Due to the strong temporal continuity present in textual information, we opted not to employ the LVC mechanism to capture local feature.
\begin{equation}
	X_j^t=F_{SA}\left(F_{GRU}\left(\mathbf{e}_j^i\right)\right)
\end{equation}
where $F_{SA}$ and $F_{GRU}$ denote the learming functions of GRU and self-attentive mechanism, respectively.

\subsubsection{Major Modality Encoder}
In WavFusion, we encode low-level audio features through the shallow transformer layer, followed by combining text and visual features through the deep transformer layer to form a comprehensive multimodal representation. We define the original transformer layer as a shallow transformer layer and the modified transformer layer as a deep transformer. The incorporation of text and vision into wav2vec 2.0 detects relevant information within the extensive pre-trained audio knowledge, thereby enhancing emotional information within the multimodal fusion representation. The low-level acoustic features $X_j^a$ extracted by the shallow transformer block are calculated as follows:
\begin{equation}
	X_j^a=F_{S T}\left(S_j^a\right)
\end{equation}
where $F_{ST}$ is the learning function of shallow transformer layers.

\begin{equation}
	X_j^{F1}=CM_{A-T}\left(X_j^a, X_j^t\right)
\end{equation}

\begin{equation}
	X_j^{F2}=CM_{A-V}\left(X_j^a, X_j^v\right)
\end{equation}	

Finally, the augmented features $X_j^{F1}$ and $X_j^{F2}$ are processed through the following gated filtering mechanism. The ratio of each channel can be dynamically defined by a learnable parameter that filters out misinformation generated during cross-modal interactions.
\begin{equation}
	P_*={sigmoid}\left(FC\left(X_j^{F1}\oplus X_j^{F1}\right)\right)
\end{equation}
\begin{equation}
	X_j^F=P_* \odot X_j^{F 1}+\left(1-P_*\right) \odot X_j^{F2}
\end{equation}

\subsubsection{Multimodal Homogeneous Feature Discrepancy Learning}

Multimodal homogeneous feature discrepancy learning has made significant progress in multimodal emotion recognition. It can optimize the modal representation ability and extract richer and more accurate emotional information by learning the relationships and differences between homogeneous features. First, we feed unfused audio features $X_j^a$, text features $X_j^t$ and visual features $X_j^v$ into a shared encoder to obtain homogeneous features. It minimizes the feature gap from different modalities and contributes to multimodal alignment.
\begin{equation}
	X_{com}^{m[i]}=S D\left(X_j^m\right), m \in(a, t, v)
\end{equation}
where $SD$ is the shared encoder learning function that consists of a simple linear layer.

In this study, we perform multimodal homogeneous feature discrepancy learning to enhance the interactions between the same emotions but different modalities, and amplify the differences between the same modalities but different emotions. We define this loss function as margin loss.
\begin{eqnarray}
	\nonumber L_{mar} = \frac{1}{M}\sum_{(i,j,k) \in M} max(0,\alpha-cos(X_{com}^{m[i],c[i]}, X_{com}^{m[j],c[j]})) \\
	+ cos(X_{com}^{m[i],c[i]}, X_{com}^{m[k],c[k]}))
\end{eqnarray}
where
$$
M=\{(i, j, k) \mid m[i] \neq m[j], m[i]=m[k], c[i]=c[j], c[i] \neq c[k]\}
$$
is the modality of the sample $i$, and the $c[i]$ is the label of sample $i$. $\cos$ denotes the cosine similarity between two feature vectors. By applying a distance margin $\alpha$, we ensure that the distance between positive samples is smaller than the distance between negative samples. Here, positive samples refer to the same emotion but different emotions, and negative samples refer to the same modality but different emotions.

Similarly, cross-entropy serves as a commonly employed loss function for optimizing model parameters and enhancing classification accuracy during training.  
\begin{equation}
	L_{ {task}}^{ {emotion}}=-\frac{1}{N_D} \sum_{j=0}^{N_D} y_j \cdot \log \hat{y_j}
\end{equation}
where $y_j$ is the true label of the sample, $\hat{y_j}$ is the prediction of the sample, and $N_D$ is the number of samples in the dataset $D$.

\begin{equation}
	L_{{total}}^{ {emotion}}=L_{ {task}}^{ {emotion}}+\lambda L_{mar}
\end{equation}
where $\lambda$ is the balance factor.

\section{Evaluation}
\subsection{Dataset}
We evaluate our proposed method on two prevalent benchmark datasets for ERC, including IEMOCAP \cite{busso2008iemocap} and MELD \cite{poria2018meld}, respectively. 
The IEMOCAP dataset consists of 12 hours of improvised and scripted audio-visual data from 10 UC theatre actors (five males and five females). The dataset is divided into five binary sessions, and each conversation is annotated with emotional information in four modalities: video, audio, transcription, and motion capture of facial movements. We evaluate our model using audio, transcribed, and video data. The dataset contains a total of 7380 data samples. E.g., happy, neutral, angry, excited, sad, and frustrated. For evaluation, we employ a five-fold cross-validation approach. The first four sessions are utilized as the training set and the validation set, and the last session is utilized as the testing set. 

The MELD dataset is derived from over 1,400 dialogues and 13,000 utterances extracted from the TV series Friends. Each utterance in the dataset is annotated with one of seven emotion labels: neutral, surprise, fear, sadness, joy, disgust, and anger. The dataset includes multimodal scenes, making it suitable for studying multimodal emotion recognition tasks. For our experiments, we utilize the predefined training/validation splits provided with the MELD dataset. This ensures consistency with existing approaches and allows for a fair comparison with other models. 

\subsection{Setting}
For text and visual modalities, we freeze the parameters in the RoBERTa and EfficientNet pre-trained models and treat them as a feature extractor. The last dimension of the text and visual features is 768 and 64. For speech modalities, we unfreeze the parameters of the deep transformer layer in the wav2vec 2.0 pre-train model. These parameters are updated during model training, while the parameters of the other layers are freezing. The last dimension of the speech features are 768 and 64.

\begin{table}[htbp]
	\centering
	\caption{The results of different methods on the IEMOCAP database.}
	\setlength{\tabcolsep}{8mm}	
	\renewcommand{\arraystretch}{1.2}
	\begin{tabular}{c|ccc}
		\toprule
		Method & ACC(\%) & WF1(\%) & Year \\
		\midrule
		DialogueTRM \cite{mao2020dialoguetrm} & 68.92 & 69.23 & 2020 \\
		HiTrans \cite{li2020hitrans}& -     & 64.5  & 2020 \\
		DialogXL \cite{shen2021dialogxl}& -     & 65.94 & 2021 \\
		MMGCN \cite{hu2021mmgcn}& -     & 66.22 & 2021 \\
		COGMEN \cite{joshi2022cogmen}& 68.2  & 67.63 & 2022 \\
		MM-DFN \cite{hu2022mm} & 68.21 & 68.18 & 2022 \\
		M2FNet \cite{chudasama2022m2fnet}& 69.69 & 69.86 & 2022 \\
		HAAN-ERC \cite{zhang2023haan}& 69.48 & 69.47 & 2023 \\
		\textbf{Ours}  & \textbf{70.53} & \textbf{70.6}  & \textbf{2024} \\
		\bottomrule
	\end{tabular}%
	\label{tab:addlabel}%
\end{table}%

\begin{table}[htbp]
	\centering
	\caption{The results of different methods on the MELD database.}
	\setlength{\tabcolsep}{8mm}	
	\renewcommand{\arraystretch}{1.2}
	\begin{tabular}{c|ccc}
		\toprule
		Method & ACC(\%) & WF1(\%) & Year \\
		\midrule
		DialogueTRM \cite{mao2020dialoguetrm}& 65.66 & 63.55 & 2020 \\
		MMGCN \cite{hu2021mmgcn}& -     & 58.65 & 2021 \\
		MM-DFN \cite{hu2022mm}& 62.49 & 59.46 & 2022 \\
		UniMSE \cite{hu2022unimse}& 65.09 & 65.51 & 2022 \\
		HAAN-ERC \cite{zhang2023haan}& 66.5  & 65.66 & 2023 \\
		\textbf{Ours}  & \textbf{66.93} & \textbf{66.1}  & \textbf{2024} \\
		\bottomrule
	\end{tabular}%
	\label{tab:addlabel}%
\end{table}%

\subsection{Comparative Analysis}
In Tables 1 and 2, we show the performance of different approaches on the IEMOCAP and MELD datasets. The evaluation metrics are Accuracy (ACC) and Weighted F1 score (WF1). On the IEMOCAP dataset, our method outperformed the state-of-the-art by 0.84\% in ACC and 0.74\% in WF1. Similarly, on the MELD dataset, our method surpassed the state-of-the-art by 0.43\% in ACC and 0.44\% in WF1. The reasons are probably twofold. Firstly, we argue that this is because most of these models do not explicitly consider redundant information in the cross-modal fusion process, but our proposed method considers these through a gated cross-modal attention mechanism. Secondly, most of them only take into account the distances of different emotion samples of the same modality, but not the distances of the same emotion samples of different modalities.

\subsection{Ablation Studies}
To verify the effectiveness of WavFusion model, we conduct ablation studies on the IEMOCAP dataset. First, we reveal the importance of each modality in this section. Specifically, when utilizing a single modality, we omitted the gated cross-modal attention and multimodal homogeneous feature discrepancy learning. The results in Table 3 illustrate that the highest accuracy and weighted average F1 scores are attained when incorporating all three modalities. Due to the complexity of emotion recognition, recognizing emotions using a single modality is challenging to meet the demands of reality. We can achieve better recognition performance by integrating multimodal information. 

\begin{table}[htbp]
	\centering
	\caption{Experiment results on the diffferent modalities.}
	\setlength{\tabcolsep}{13mm}
	\renewcommand{\arraystretch}{1.2}	
	\begin{tabular}{c|cc}
		\toprule
		Modality & ACC(\%)   & WF1(\%) \\
		\midrule
		A     & 66.06 & 65.59 \\
		T     & 58.74 & 58.63 \\
		V     & 29.88 & 26.31 \\
		A+T   & 67.75 & 67.45 \\
		A+V   & 66.33 & 64.14 \\
		A+V+T & 70.53 & 70.6 \\
		\bottomrule
	\end{tabular}%
	\label{tab:addlabel}%
\end{table}%

Additionally, we introduce LVC blocks to capture local information related to visual features. To assess the significance of LVC blocks, we conducted an experiment where we omitted the LVC blocks from the model, thus failing to capture local information about visual features. From Table 4, we observe that the model with the LVC block outperforms the model without the LVC block. The inclusion of LVC blocks improves ACC by 0.63\% and the WF1 by 0.76\%. The experiment demonstrates that the LVC blocks are beneficial for capturing relevant contextual details and spatial dependencies.

\begin{table}[t]
	\centering
	\caption{Experiment results on LVC BLOCK.}
	\setlength{\tabcolsep}{12mm}
	\renewcommand{\arraystretch}{1.2}	
	\begin{tabular}{c|cc}
		\toprule
		Models & ACC(\%)   & WF1(\%) \\
		\midrule
		w/o LVC block     & 69.90 & 69.84 \\
		w/ LVC block     & 70.53 & 70.60 \\
		\bottomrule
	\end{tabular}%
	\label{tab:addlabel}%
\end{table}%

We also investigate the impact of multimodal homogeneous feature discrepancy learning in our framework. In this work, we assigned weights to the balance factor $\lambda$ for margin loss and observed its effects across various weight values. The corresponding results are presented in Table 5. The results indicate that the optimal performance on the IEMOCAP dataset is achieved. The model shows a significant improvement by 2.64\% and WF1 by 2.94\% compared to the absence of margin loss ($\lambda$ = 0). This demonstrates the effectiveness of multimodal homogeneous feature difference learning in enhancing the model's capacity to discern emotions across diverse modalities. However, we also observe that the performance deteriorates when the balance factor is excessively large ($\lambda$ = 10). This suggests that an excessive emphasis on margin loss might have a detrimental effect on the original classification task.

\begin{table}[t]
	\centering
	\caption{Experiment results on the diffferent $\lambda$.}
	\setlength{\tabcolsep}{14mm}	
	\renewcommand{\arraystretch}{1.2}
	\begin{tabular}{c|cc}
		\toprule
		$\lambda$	& ACC(\%)   & WF1(\%) \\
		\midrule
		0     & 67.89 & 67.66 \\
		0.01  & 68.63 & 68.39 \\
		0.1   & 69.11 & 68.96 \\
		1     & 70.53 & 70.6 \\
		10    & 64.43 & 64.19 \\
		\bottomrule
	\end{tabular}%
	\label{tab:addlabel}%
\end{table}%
\vspace{10mm}
\begin{table}[t]
	\centering
	\caption{Experiment results on the transformed layers in wav2vec 2.0.}
	\setlength{\tabcolsep}{3mm}	
	\begin{tabular}{c|cccc}
		\toprule
		\multirow{2}[2]{*}{method} & \multicolumn{1}{c}{\multirow{2}[2]{*}{Shallow transformer}} & \multicolumn{1}{c}{\multirow{2}[2]{*}{Deep transformer}} & \multirow{2}[2]{*}{ACC(\%)} & \multirow{2}[2]{*}{WF1(\%)} \\
		&       &       &       &  \\
		\midrule
		concat & 12    & 0     & 66.67 & 66.78 \\
		\midrule
		\multirow{4}[2]{*}{Attention} & 11    & 1     & 68.61 & 68.55 \\
		& 10    & 2     & 68.54 & 68.32 \\
		& 9     & 3     & 70.53 & 70.6 \\
		& 8     & 4     & 69.29 & 69.06 \\
		\bottomrule
	\end{tabular}%
	\label{tab:addlabel}%
\end{table}%

We also observe the effect of gated cross-modal attention mechanism in the proposed framework. In our experiments, we define the original transformer layer the shallow transformer and the modified transformer layer as deep transformer, and observe their effect on the different numbers. In the first line, we omit the proposed gated cross-modal attention mechanism and solely conduct a basic concatenation of the three modal features at the last dimension. The corresponding results are shown in Table 6 where it is observed that 9 shallow transformer layers, 3 deep transformer layers yield the optimal performance for the IEMOCAP dataset. Moreover, from the first and second lines, we can discern the significance of the gated cross-modal attention mechanism for fusion.

\section{Conclusion}
In this paper, we propose a novel SER approach, which is designed a gated cross-modal attention alternative to self-attention in the wav2vec 2.0 pre-trained model to dynamically fuse features from different modalities. Additionally, we introduce a novel LVC block to efficiently capture the local information of visual features. The model can more effectively utilize the spatial characteristics of visual data, resulting in more comprehensive representations. Finally, we design the concept of multimodal homogeneous feature discrepancy learning, which helps the model to effectively learn and distinguish representations of the same modalities but different emotions. The effectiveness of the proposed model is demonstrated on the IEMOCAP and MELD datasets. The results show promising performance compared to state-of-the-art methods. In the future, we plan to utilize the leveraging large amounts of unlabeled audio and video data available to recognize the different emotion.

\begin{credits}
	\subsubsection{\ackname} This work was supported in part by the Natural Science Foundation of the Higher Education Institutions of Anhui Province under Grant Nos. 2024AH050018 and KJ2021A0486, Excellent Research and Innovation Team of Universities at Anhui Province under Grant Nos. 2024AH010001 and 2023AH010008, and Science Research Fund of Anhui University of Finance and Economics under Grant No. ACKYB23016.

\end{credits}

\end{document}